\documentclass[nohyper,letterpaper,12pt]{JHEP3}

\usepackage{amsfonts,amssymb,amsmath,graphicx}

\title{Moduli Dynamics of AdS$_\mathbf{3}$ Strings}

\author{Antal Jevicki, Kewang Jin \\ Department of Physics, Brown University, Box 1843, Providence, RI 02912, USA \\ E-mail: \email{antal@het.brown.edu, jin@het.brown.edu}}

\abstract{We construct a general class of solutions for a classical string in $AdS_3$ spacetime. The construction is based on a Pohlmeyer type reduction, with the sinh-Gordon model providing the general N-soliton solutions. The corresponding exact spiky string configurations are then reconstructed through the inverse scattering method. It is shown that the string moduli are determined entirely by those of the solitons.}


\preprint{\tt{BROWN-HET-1578}}

\begin{document}

\section{Introduction}

It is of definite relevance to understand fully the semiclassical dynamics of strings moving in the AdS background. Through the AdS/CFT correspondence this provides the strong coupling limit of higher spin states of ${\cal N}=4$ Super Yang-Mills theory and offers further insight into the general Bethe ansatz solution. For states of Super Yang-Mills theory given by single trace operators with insertions of (covariant) derivatives the relevant string configurations were identified by Gubser, Klebanov and Polyakov \cite{Gubser:2002tv} as rotating folded strings. Kruczenski gave the extension to regular N-spike configurations in \cite{Kruczenski:2004wg}. These configurations in some sense represent the ground state configurations of the theory and it is of definite relevance to construct (and understand the dynamics) of more general non-static solutions \cite{Dorey:2008zy, Alday:2007hr}. Ultimately one has the goal of formulating a complete dynamical picture of the moduli of strings moving in $AdS \times S$ spacetimes.

In this paper we continue our previous work \cite{Jevicki:2007aa} and present the construction of a general set of time dependent classical string solutions in $AdS_3$. In \cite{Jevicki:2007aa} we have developed the inverse scattering method for that purpose following the well known strategy of reducing the string sigma models \cite{Pohlmeyer,Barbashov:1982nc} to integrable scalar field theory systems \cite{de Vega}. For $AdS_3$ this was given by the sinh-Gordon theory which possesses both regular and singular solutions. We have in \cite{Jevicki:2007aa} introduced a picture where the soliton solutions are seen to be in a one to one correspondence with the spike solutions of the AdS string. This we have demonstrated in the case of Kruczenski's static N-spike configuration and also in the time dependent cases corresponding to one and two soliton solutions. In the present work we sharpen this picture further and present the construction of general (spiky) string solutions associated with the most general N-soliton configurations on an infinite line. Our general solution is given in terms of two arbitrary functions ($u(\bar{z}),v(z)$) representing the conformal frame and a discrete set of (collective) coordinates representing the solitons. After fixing the conformal frame these later give the moduli space for the string.

Based on this construction we are led to a picture where singularities of the field theory configuration translate through a direct map into spikes of the AdS string. Solitons in field theory are described in general by their moduli (collective coordinates) representing their locations and momenta. An identical particle-like interpretation turns out to hold for singularities (this has been argued in a series of papers \cite{Dzhordzhadze:1979pp,Bowtell}). In our string theory representation the singular soliton coordinates map directly into coordinates associated with spike configurations. This mapping then provides a complete dynamical description of moduli associated with the general string solutions. 

It is useful to draw a parallel with an analogous construction found for the case of strings moving on $R \times S^n$. In the $n=2$ case one had the association (due to Hofman and Maldacena \cite{Hofman:2006xt}) between solitons of sine-Gordon and magnons of the string (see also \cite{Chen:2006gea} for generalizations). In \cite{Aniceto:2008pc} the moduli space description was presented, given by the N-body problem of Calogero (or Ruijsenaars-Schneider type \cite{Ruijsenaars}). The present paper implies an analogous description in the case of $AdS_3$. In the 2D non-critical string case one has the long folded strings of \cite{Maldacena:2005hi} (see also \cite{Bars:1994qm}).

With the present discussion being on the infinite line, one has the important technical future extension to the periodic circle. This work is in progress. Nevertheless the study in the non-compact setting exhibits many of the main features that are expected in the periodic case also. In particular the association of singularities with spikes, and their moduli spaces are to appear very analogously. 

The content of the paper is as follows. In section 2 we clarify the singularity-spike map in the flat spacetime with a detailed study. In section 3 we shortly summarize the basis of the method and present the explicit construction of $AdS_3$ string solutions associated with the general N-soliton solutions. Also we discuss the associated moduli space N-body dynamics. The appendices are reserved for the more technical details.

\section{Spiky strings in flat spacetime}

\subsection{Pohlmeyer reduction and the Liouville equation}

It is useful to begin by summarizing the construction of the much simpler case of flat three dimensional string representing
the $R \to \infty$ limit of the AdS spacetime. In this case the reduced theory is given by the Liouville equation \cite{Barbashov:1982nc,Dzhordzhadze:1979pp} whose general solutions are explicitly given.

The conformal gauge equations of motion for strings in flat spacetime are
\begin{equation}
\partial_+ \partial_- X=0,
\label{eqn21}
\end{equation}
as well as the Virasoro constraints
\begin{equation}
(\partial_+ X)^2=(\partial_- X)^2=0,
\label{eqn22}
\end{equation}
where $X^\mu=(X^0,X^1,X^2)$ with the flat metric $\eta_{\mu\nu}=\{-1,1,1\}$. Here we follow the notation $\sigma^\pm=\tau \pm \sigma$ so that $\partial_\pm=(\partial_\tau \pm \partial_\sigma)/2$, where $\tau,\sigma$ are the Minkowski worldsheet coordinates of the string. Defining the scalar field \footnote{A minus sign is added to the definition for convention.}
\begin{equation}
\alpha(\sigma^+,\sigma^-) \equiv -\ln[\partial_+ X \cdot \partial_- X],
\end{equation}
we find the equation of motion for $\alpha$ to be
\begin{equation}
\partial_+ \partial_- \alpha(\sigma^+,\sigma^-)-u(\sigma^+)v(\sigma^-) e^{\alpha}=0,
\label{eqn24}
\end{equation}
where $u(\sigma^+)$ and $v(\sigma^-)$ are two arbitrary functions. The general solution to the Liouville equation (\ref{eqn24}) reads
\begin{equation}
\alpha=\ln\Bigl[{2 \over u(\sigma^+)v(\sigma^-)}{f'(\sigma^+)g'(\sigma^-) \over [f(\sigma^+)+g(\sigma^-)]^2}\Bigr].
\end{equation}

In order to understand the most general form of the string solution we note that in the conformal gauge one still has a residual 
symmetry.  Both the equations of motion (\ref{eqn21}) and the Virasoro constraints (\ref{eqn22}) are invariant with respect to the conformal transformations $\sigma^+ \rightarrow f(\sigma^+),\sigma^- \rightarrow g(\sigma^-)$. We can, without loss of generality, specify the conformal frame by the following conditions
\begin{equation}
\dot{X}_{,+}^2=u^2(\sigma^+), \qquad \dot{X}_{,-}^2=v^2(\sigma^-).
\label{eqn26}
\end{equation}
The general solution to the equations of motion (\ref{eqn21}) satisfying the Virasoro constraints (\ref{eqn22}) is now constructed following \cite{Barbashov:1982nc} as
\begin{equation}
X^\mu(\sigma^+,\sigma^-)=\psi_+^\mu (\sigma^+)+\psi_-^\mu(\sigma^-),
\label{eqn27}
\end{equation}
where $\psi_+^{'\mu}$ and $\psi_-^{'\mu}$ being the isotropic vectors
\begin{equation}
(\psi_\pm^\prime)^2=0.
\label{eqn28}
\end{equation}
The prime implies the differentiation with respect to the function argument. Substituting (\ref{eqn27}) into ({\ref{eqn26}), we obtain one more condition on $\psi_\pm^\mu$
\begin{equation}
(\psi_+^{''})^2=u^2(\sigma^+), \qquad (\psi_-^{''})^2=v^2(\sigma^-).
\label{eqn29}
\end{equation}
The conditions (\ref{eqn28}) and (\ref{eqn29}) can easily be satisfied by expanding the vectors $\psi_\pm^{'\mu}$ in a special basis. In the case of three dimensions, we choose the basis
\begin{equation}
e_1={1 \over \sqrt{2}}\begin{pmatrix} 1 \cr 1 \cr 0 \end{pmatrix}, \quad e_2={1 \over \sqrt{2}}\begin{pmatrix} 1 \cr -1 \cr 0 \end{pmatrix}, \quad e_3=\begin{pmatrix} 0 \cr 0 \cr 1 \end{pmatrix}.
\end{equation}
The expansion for $\psi_\pm^\prime(\sigma^\pm)$ in this basis can be written as
\begin{eqnarray}
\psi_+^\prime (\sigma^+)&=&+{u(\sigma^+) \over f'(\sigma^+)}\Bigl[e_1+{1 \over 2}f^2(\sigma^+) e_2+f(\sigma^+) e_3 \Bigr], \label{eqn211} \\
\psi_-^\prime (\sigma^-)&=&-{v(\sigma^-) \over g'(\sigma^-)}\Bigl[e_1+{1 \over 2}g^2(\sigma^-) e_2-g(\sigma^-) e_3 \Bigr]. \label{eqn212}
\end{eqnarray}
The spikes are located at $(X')^2=-2 e^{-\alpha}=0$, which generates the condition
\begin{equation}
f(\sigma^+)+g(\sigma^-)=0.
\label{eqn213}
\end{equation}
As an example, take the spiky string solution in \cite{Kruczenski:2004wg}, we have
\begin{equation}
\psi_+(\sigma^+)=\begin{pmatrix} \lambda (n-1) \sigma^+ \cr \lambda \cos((n-1)\sigma^+) \cr \lambda \sin((n-1)\sigma^+) \end{pmatrix}, \quad \psi_-(\sigma^-)=\begin{pmatrix} \lambda (n-1)\sigma^- \cr \lambda (n-1)\cos(\sigma^-) \cr \lambda (n-1) \sin(\sigma^-) \end{pmatrix},
\end{equation}
so that
\begin{equation}
u=\lambda(n-1)^2, \qquad v=\lambda(n-1).
\end{equation}
Using the representation (\ref{eqn211}) and (\ref{eqn212}), we find
\begin{equation}
f(\sigma^+)=\sqrt{2}\cot{(n-1)\sigma^+ \over 2}, \qquad g(\sigma^-)=-\sqrt{2}\cot{\sigma^- \over 2}.
\end{equation}
The condition (\ref{eqn213}) generates the locations of spikes
\begin{equation}
(n-1)\sigma^+-\sigma^-=2\pi m,
\end{equation}
which of course agrees with \cite{Kruczenski:2004wg}.

\subsection{Dynamics of singularities and spikes}

As we have understood in our previous work, spikes in the string configuration are associated with locations of solitons (singularities) in the scalar field solution. For the Liouville theory a detailed study of the dynamics of singularities was given in \cite{Dzhordzhadze:1979pp}. They are determined by the equation (\ref{eqn213}) giving a description of the world lines of dynamical particles. This interpretation is suggested by the time-like nature of the singularity lines and the fact that each line is characterized by the initial data $\sigma_j^0$ and $v_j$. Explicitly, if $\sigma_j(\tau)$ is the equation of the j-th singularity line, then we have $\sigma_j(0)=\sigma_j^0,\dot{\sigma}_j(0)=v_j$.

To summarize the discussion of \cite{Dzhordzhadze:1979pp}, one starts with arbitrary functions $f(\sigma^+)$ with $N_A$ singularities and $g(\sigma^-)$ with $N_B$ singularities
\begin{equation}
f(\sigma^+)=\sum_{j=1}^{N_A}{c_j \over y_j-\sigma^+}, \qquad g(\sigma^-)=\sum_{j=1}^{N_B}{d_j \over z_j-\sigma^-},
\label{eqn218}
\end{equation}
where $c_j,d_j,y_j,z_j$ are constants. The number of singularities determined by (\ref{eqn213}) is $N=N_A+N_B-1$, which we denote by $\sigma^\pm_i(\tau)$, where $\tau$ is the variable parameterizing the lines, and for convenience we assume that
\begin{equation}
\sigma^\pm_i(0)=\pm \sigma_i^0, \qquad \dot{\sigma}^\pm_i(0)=1 \pm v_i, \qquad i=1,2,\cdots,N.
\label{inidata}
\end{equation}
Thus, we obtain the system
\begin{equation}
\sum_{j=1}^{N_A}{c_j \over y_j-\sigma^+_i}+\sum_{j=1}^{N_B}{d_j \over z_j-\sigma^-_i}=0, \qquad i=1,2,\cdots,N.
\label{diff0}
\end{equation}
The constants $c_j,d_j,y_j,z_j$ in this system must be determined from the initial data of the Liouville field. Here, this can be seen directly. We differentiate (\ref{diff0}) with respect to $\tau$,
\begin{equation}
\dot{\sigma}^+_i \sum_{j=1}^{N_A}{c_j \over (y_j-\sigma^+_i)^2}+\dot{\sigma}^-_i \sum_{j=1}^{N_B}{d_j \over (z_j-\sigma^-_i)^2}=0, \qquad i=1,2,\cdots,N.
\label{diff1}
\end{equation}
Setting now $\tau=0$ in (\ref{diff0}) and (\ref{diff1}) and using (\ref{inidata}), we obtain a system of $2N=2(N_A+N_B)-2$ equations for determining the $2(N_A+N_B)$ constants $c_j,d_j,y_j,z_j$. The remaining two-parameter arbitrariness exactly coincides with the arbitrariness of the restricted Bianchi transformation.

To obtain the dynamical equations of motion whose solutions are to be these singularities lines, we can differentiate further (\ref{diff1}) with respect to $\tau$. These really are the equations of motion, since the constants $c_j,d_j,y_j,z_j$ can be expressed in accordance with (\ref{diff0}) and (\ref{diff1}) in terms of $\sigma^+_i,\dot{\sigma}^+_i,\sigma^-_i,\dot{\sigma}^-_i$ not only for $\tau=0$ but also at any time $\tau$, in particular the time at which we consider the system equations of motion.

Turning now to the string solution, it was given generally by (\ref{eqn211}, \ref{eqn212}) with two arbitrary functions $f(\sigma^+)$ and $g(\sigma^-)$ and the functions $u(\sigma^+)$ and $v(\sigma^-)$ representing the conformal frame. After fixing these, one can in principle integrate (\ref{eqn211}, \ref{eqn212}) to determine the string solution. A particular interesting class of these functions are those with singularities described above. These singularities in field theory will translate to spikes in string theory. To exemplify this connection, we will describe the simplest cases with one and two singularities, i.e., one and two spikes.

In the case of one singularity, set $N=N_A=N_B=1$ in (\ref{eqn218}), using the initial data (\ref{inidata}), we can solve for the constants
\begin{equation}
c_1=-d_1{1+v_1 \over 1-v_1}, \qquad y_1={2\sigma_1^0 \over 1-v_1}+z_1{1+v_1 \over 1-v_1}.
\end{equation}
so that the trajectory of the singularity is
\begin{equation}
\sigma_1(\tau)=\sigma_1^0+v_1 \tau.
\end{equation}
By integrating (\ref{eqn211}) and (\ref{eqn212}), we get the string solution
\begin{eqnarray}
X^0&=&{u \over \sqrt{2}d_1\tilde{v}_1}\Bigl({1 \over 3}(\tilde{\sigma}^+)^3+{1 \over 2}d_1^2 \tilde{v}_1^2 \tilde{\sigma}^+ \Bigr)+{v \over \sqrt{2}d_1}\Bigl({1 \over 3}(\tilde{\sigma}^-)^3+{1 \over 2}d_1^2 \tilde{\sigma}^- \Bigr), \\
X^1&=&{u \over \sqrt{2}d_1\tilde{v}_1}\Bigl({1 \over 3}(\tilde{\sigma}^+)^3-{1 \over 2}d_1^2 \tilde{v}_1^2 \tilde{\sigma}^+ \Bigr)+{v \over \sqrt{2}d_1}\Bigl({1 \over 3}(\tilde{\sigma}^-)^3-{1 \over 2}d_1^2 \tilde{\sigma}^- \Bigr), \\
X^2&=&{u \over 2}(\tilde{\sigma}^+)^2+{v \over 2}(\tilde{\sigma}^-)^2,
\end{eqnarray}
where, for simplicity, $u$ and $v$ are chosen to be constants and the redefinitions
\begin{equation}
\tilde{\sigma}^+ \equiv \sigma^+-{2\sigma_1^0 \over 1-v_1}-z_1{1+v_1 \over 1-v_1}, \qquad \tilde{\sigma}^- \equiv \sigma^--z_1, \qquad \tilde{v}_1 \equiv {1+v_1 \over 1-v_1}.
\end{equation}
We can generalize the above case to the `periodic' case with the identity
\begin{equation}
\cot{x \over 2}=\sum_{n=-\infty}^\infty {2 \over x+2 \pi n},
\end{equation}
and find
\begin{equation}
f(\sigma^+)={d_1 \tilde{v}_1 \over 2}\cot{\tilde{\sigma}^+ \over 2}, \qquad g(\sigma^-)=-{d_1 \over 2}\cot{\tilde{\sigma}^- \over 2}.
\end{equation}
After the integration, the string solution is found to be
\begin{eqnarray}
X^0&=&-{\sqrt{2}u \over d_1 \tilde{v}_1}\Bigl( (\tilde{\sigma}^+-\sin\tilde{\sigma}^+)+{d_1^2 \tilde{v}_1^2 \over 8}(\tilde{\sigma}^++\sin\tilde{\sigma}^+)\Bigr) \cr
& &-{\sqrt{2}v \over d_1}\Bigl( (\tilde{\sigma}^--\sin\tilde{\sigma}^-)+{d_1^2 \over 8}(\tilde{\sigma}^-+\sin\tilde{\sigma}^-) \Bigr), \label{eqn230} \\
X^1&=&-{\sqrt{2}u \over d_1 \tilde{v}_1}\Bigl( (\tilde{\sigma}^+-\sin\tilde{\sigma}^+)-{d_1^2 \tilde{v}_1^2 \over 8}(\tilde{\sigma}^++\sin\tilde{\sigma}^+)\Bigr) \cr
& &-{\sqrt{2}v \over d_1}\Bigl( (\tilde{\sigma}^--\sin\tilde{\sigma}^-)-{d_1^2 \over 8}(\tilde{\sigma}^-+\sin\tilde{\sigma}^-) \Bigr), \label{eqn231} \\
X^2&=&u\cos\tilde{\sigma}^++v\cos\tilde{\sigma}^-. \label{eqn232}
\end{eqnarray}
It is interesting to notice the special case where
\begin{equation}
v_1=0, \qquad \tilde{v}_1=1, \qquad d_1=2\sqrt{2},
\end{equation}
the string solution (\ref{eqn230}$-$\ref{eqn232}) reduces to the spiky strings in \cite{Kruczenski:2004wg} with two spikes ($n=2$) if we identify
\begin{equation}
\tilde{\sigma}^+={\pi \over 2}-\sigma_+, \qquad \tilde{\sigma}^-={\pi \over 2}-\sigma_-, \qquad u=\lambda, \qquad v=\lambda.
\end{equation}

\section{Spiky strings in AdS spacetime}

We will now turn to the general construction of the $AdS_3$ case. We will begin by giving a summary of the method used in our previous work. In general, sting equations in $AdS_d$ spacetime (in conformal gauge) are described by the non-compact nonlinear sigma model on $SO(d-1,2)$. Defining the $AdS_d$ space as $Y^2=-Y_{-1}^2-Y_0^2+Y_1^2+\cdots+Y_{d-1}^2=-1$, the action reads
\begin{equation}
S={\sqrt{\lambda} \over 2\pi}\int d\tau d\sigma \Bigl(\partial Y \cdot \partial Y + \lambda (\sigma ,\tau)(Y \cdot Y+1)\Bigr),
\end{equation}
where $\tau,\sigma$ are the Minkowski worldsheet coordinates, the equations of motion are
\begin{equation}
\partial \bar{\partial} Y-(\partial Y \cdot \bar{\partial} Y) Y=0,
\end{equation}
with $z=(\sigma-\tau)/2,\bar{z}=(\sigma+\tau)/2$ and $\partial=\partial_\sigma-\partial_\tau,\bar{\partial}=\partial_\sigma+\partial_\tau$. In addition to guarantee the conformal gauge we have to impose the Virasoro conditions
\begin{equation}
\partial Y \cdot \partial Y = \bar{\partial} Y \cdot \bar{\partial} Y=0.
\end{equation}

It was demonstrated a number of years ago (by Pohlmeyer \cite{Pohlmeyer}) that nonlinear sigma models subject to Virasoro type constraints can be reduced to known, integrable field equations of sine-Gordon (or Toda) type. This reduction is accomplished by concentrating on $SO(d-1,2)$ invariant sub-dynamics of the sigma model. The steps of the reduction were well described in \cite{de Vega,Miramontes:2008wt} and consist in the following. One starts by identifying first an appropriate set of basis vectors for the string coordinates. For $AdS_3$, the basis can be chosen as
\begin{equation}
e_i=(Y,\bar{\partial} Y,\partial Y,B_4), \qquad i=1,\cdots,4,
\end{equation}
where $B_4$ is a fourth orthonormal vector, satisfying $B_4 \cdot B_4=1,B_4 \cdot Y=B_4 \cdot \partial Y=B_4 \cdot \bar{\partial} Y=0$. The reduced (invariant) scalar field is introduced through a scalar product
\begin{equation}
\alpha(z,\bar{z}) \equiv \ln \bigl[\partial Y \cdot \bar{\partial} Y\bigr],
\end{equation}
and one proceeds to derive the equation of motion for $\alpha$ which reads
\begin{equation}
\partial \bar{\partial} \alpha(z,\bar{z})-e^\alpha-u(\bar{z})v(z)e^{-\alpha}=0,
\end{equation}
where $u(\bar{z})$ and $v(z)$ are two additional (invariant) scalar fields given by $u=B_4 \cdot \bar{\partial}^2 Y,v=B_4 \cdot \partial^2 Y$. They are found to obey the equations $\partial u=\bar{\partial} v=0$ and now the closed set of equations defines the generalized sinh-Gordon model (see \cite{Jevicki:2007aa} for more details).

\subsection{Gauge transformation}

We had the fact that the classical strings in $AdS_3$ can be reduced to a generalized sinh-Gordon model coupled to two arbitrary functions $u(\bar{z})$ and $v(z)$ which together represent a free scalar field. These functions are central to the string theory interpretation of the sinh-Gordon equation, they represent the freedom of performing general conformal transformations which are the symmetry of the conformal gauge string. In the system of coupled equations describing the Lax pair they can be transformed by a combination of conformal and gauge transformations as we now demonstrate.

We start with a redefinition (shift) of the sinh-Gordon field
\begin{equation}
\alpha(z,\bar{z})=\hat{\alpha}(z,\bar{z})+\ln \sqrt{-u(\bar{z})v(z)},
\end{equation}
and then the new field $\hat{\alpha}$ satisfies the equation
\begin{equation}
\partial \bar{\partial} \hat{\alpha}(z,\bar{z})-\sqrt{-4u(\bar{z})v(z)}\sinh \hat{\alpha}(z,\bar{z})=0.
\end{equation}
Now after a (conformal) change of variables
\begin{equation}
\bar{z}'=\int \sqrt{2u(\bar{z})}d \bar{z}, \qquad z'=\int \sqrt{-2v(z)}d z,
\label{eqn39}
\end{equation}
we obtain the standard form of the sinh-Gordon equation
\begin{equation}
\partial' \bar{\partial}' \hat{\alpha}(z',\bar{z}')-\sinh\hat{\alpha}(z',\bar{z}')=0.
\end{equation}
Omitting the primes, we can write down the two Lax pairs in terms of the new variables
\begin{eqnarray}
\bar{\partial} \phi(z,\bar{z})&=&A_1 \phi(z,\bar{z}), \qquad \bar{\partial} \psi(z,\bar{z})=B_1 \psi(z,\bar{z}), \\
\partial \phi(z,\bar{z})&=&A_2 \phi(z,\bar{z}), \qquad \partial \psi(z,\bar{z})=B_2 \psi(z,\bar{z}),
\end{eqnarray}
where the matrices are given by
\begin{eqnarray}
A_1&=&{1 \over 4}\begin{pmatrix} -i\lambda c_1^+ & i\bar{\partial}\hat{\alpha}+{i \over 2}{u'(\bar{z}) \over u(\bar{z})}-\lambda c_1^- \cr -i\bar{\partial}\hat{\alpha}-{i \over 2}{u'(\bar{z}) \over u(\bar{z})}-\lambda c_1^- & i\lambda c_1^+ \end{pmatrix}, \label{eqn313} \\
A_2&=&{1 \over 4}\begin{pmatrix} i{1 \over \lambda} c_2^+ & -i\partial\hat{\alpha}-{i \over 2}{v'(z) \over v(z)}-{1 \over \lambda} c_2^- \cr i\partial\hat{\alpha}+{i \over 2}{v'(z) \over v(z)}-{1 \over \lambda} c_2^- & -i{1 \over \lambda} c_2^+ \end{pmatrix}, \label{eqn314} \\
B_1&=&{1 \over 4}\begin{pmatrix} -i\lambda c_1^- & i\bar{\partial}\hat{\alpha}+{i \over 2}{u'(\bar{z}) \over u(\bar{z})}-\lambda c_1^+ \cr -i\bar{\partial}\hat{\alpha}-{i \over 2}{u'(\bar{z}) \over u(\bar{z})}-\lambda c_1^+ & i\lambda c_1^- \end{pmatrix}, \label{eqn315} \\
B_2&=&{1 \over 4}\begin{pmatrix} i{1 \over \lambda} c_2^- & -i\partial\hat{\alpha}-{i \over 2}{v'(z) \over v(z)}-{1 \over \lambda} c_2^+ \cr i\partial\hat{\alpha}+{i \over 2}{v'(z) \over v(z)}-{1 \over \lambda} c_2^+ & -i{1 \over \lambda} c_2^- \end{pmatrix}, \label{eqn316}
\end{eqnarray}
with definitions for simpler expressions
\begin{eqnarray}
c_1^+& \equiv &\sqrt[4]{u \over -v}e^{-{1 \over 2}\hat{\alpha}}+\sqrt[4]{-v \over u}e^{{1 \over 2}\hat{\alpha}}, \qquad c_1^- \equiv \sqrt[4]{u \over -v}e^{-{1 \over 2}\hat{\alpha}}-\sqrt[4]{-v \over u}e^{{1 \over 2}\hat{\alpha}}, \\
c_2^+& \equiv &\sqrt[4]{-v \over u}e^{-{1 \over 2}\hat{\alpha}}+\sqrt[4]{u \over -v}e^{{1 \over 2}\hat{\alpha}}, \qquad c_2^- \equiv \sqrt[4]{-v \over u}e^{-{1 \over 2}\hat{\alpha}}-\sqrt[4]{u \over -v}e^{{1 \over 2}\hat{\alpha}}.
\end{eqnarray}
Here we introduced the spectral parameter by rescaling
\begin{equation}
z \rightarrow \lambda z, \qquad \bar{z} \rightarrow {1 \over \lambda}\bar{z},
\end{equation}
which is the standard way of introducing the spectral parameter in the Lax formulation.

Our next task is to establish a relationship between the Lax pairs (\ref{eqn313}$-$\ref{eqn316}) found in the sigma model and the standard Lax pair of the sinh-Gordon theory
\begin{equation}
U=\begin{pmatrix} -i\zeta \quad {1 \over 2}\bar{\partial}\hat{u} \cr {1 \over 2}\bar{\partial}\hat{u} \quad i\zeta \end{pmatrix}, \qquad V={i \over 4\zeta} \begin{pmatrix} \cosh \hat{u} \quad -\sinh \hat{u} \cr \sinh \hat{u} \quad -\cosh \hat{u} \end{pmatrix},
\label{lax}
\end{equation}
which satisfies the Dirac equations
\begin{equation}
\bar{\partial} \varphi(z,\bar{z})=U \varphi(z,\bar{z}), \qquad \partial \varphi(z,\bar{z})=V \varphi(z,\bar{z}).
\label{laxeqn}
\end{equation}
Here $\zeta$ is the spectral parameter and the sinh-Gordon field $\hat{u}$ satisfies the equation
\begin{equation}
\partial \bar{\partial} \hat{u}(z,\bar{z}) -\sinh \hat{u}(z,\bar{z})=0.
\label{eqn322}
\end{equation}
Defining the gauge transformation as
\begin{equation}
A_1=g_A^{-1}(U-\bar{\partial})g_A, \qquad A_2=g_A^{-1}(V-\partial)g_A,
\end{equation}
for the $A$ matrices, the transformation matrix is found to be
\begin{equation}
g_A={\sqrt{i} \over 2}\begin{pmatrix} i\Bigl( \sqrt[8]{u \over -v} e^{-{1 \over 4}\hat{\alpha}}-\sqrt[8]{-v \over u} e^{{1 \over 4}\hat{\alpha}}\Bigr) \quad \Bigl( \sqrt[8]{u \over -v} e^{-{1 \over 4}\hat{\alpha}}+\sqrt[8]{-v \over u} e^{{1 \over 4}\hat{\alpha}}\Bigr) \cr i\Bigl( \sqrt[8]{u \over -v} e^{-{1 \over 4}\hat{\alpha}}+\sqrt[8]{-v \over u} e^{{1 \over 4}\hat{\alpha}}\Bigr) \quad \Bigl( \sqrt[8]{u \over -v} e^{-{1 \over 4}\hat{\alpha}}-\sqrt[8]{-v \over u} e^{{1 \over 4}\hat{\alpha}}\Bigr) \end{pmatrix},
\label{gaugeA}
\end{equation}
with the identification $\hat{u}=-\hat{\alpha},\lambda=-2\zeta.$ Similarly, for the $B$ matrices, we find
\begin{equation}
g_B=-{i \over 2}\begin{pmatrix} i\Bigl( \sqrt[8]{u \over -v} e^{-{1 \over 4}\hat{\alpha}}+i\sqrt[8]{-v \over u} e^{{1 \over 4}\hat{\alpha}}\Bigr) \quad \Bigl( \sqrt[8]{u \over -v} e^{-{1 \over 4}\hat{\alpha}}-i\sqrt[8]{-v \over u} e^{{1 \over 4}\hat{\alpha}}\Bigr) \cr i\Bigl( \sqrt[8]{u \over -v} e^{-{1 \over 4}\hat{\alpha}}-i\sqrt[8]{-v \over u} e^{{1 \over 4}\hat{\alpha}}\Bigr) \quad \Bigl( \sqrt[8]{u \over -v} e^{-{1 \over 4}\hat{\alpha}}+i\sqrt[8]{-v \over u} e^{{1 \over 4}\hat{\alpha}}\Bigr) \end{pmatrix},
\label{gaugeB}
\end{equation}
with the identification $\hat{u}=-\hat{\alpha},\lambda=-2i\zeta.$

\subsection{Sinh-Gordon soliton dynamics}

We now review the sinh-Gordon solutions and discuss their dynamics. In contrast to the sine-Gordon case the solutions in question will have singularities at the locations of solitons. In fact we use the word `soliton' loosely, these solutions do not carry topological charges. The existence of singularities however allows for a similar physical picture, with the number of singularities (much like the topological number) being conserved by dynamics. As we will discuss the singularities provide a particle (and in the string context a spike) interpretation of the solutions.  

The general solution to the sinh-Gordon equation (\ref{eqn322}) with N solitons can be obtained using the inverse scattering method \cite{AKNS} (see appendix A for details)
\begin{equation}
\hat{u}(z,\bar{z})=\sinh^{-1} \Bigl[{4\zeta \over i}{\partial(\varphi_1 \varphi_2) \over (\varphi_1)^2-(\varphi_2)^2}\Bigr], \label{eqn326}
\end{equation}
where the components of spinor $\varphi$ are
\begin{eqnarray}
\varphi_1(\zeta,z,\bar{z})&=&-\Bigl(\sum_{j,l=1}^N {\lambda_j \over \zeta+\zeta_j}(1-A)_{jl}^{-1}\lambda_l\Bigr)e^{i \zeta \bar{z}-i z / 4\zeta}, \label{eqn327} \\
\varphi_2(\zeta,z,\bar{z})&=&\Bigl(1+\sum_{j,l,k=1}^N {\lambda_j \over \zeta+\zeta_j}{\lambda_j \lambda_l \over \zeta_j+\zeta_l}(1-A)_{lk}^{-1} \lambda_k \Bigr)e^{i \zeta \bar{z}-i z / 4\zeta}, \label{eqn328}
\end{eqnarray}
with the definitions
\begin{equation}
A_{ij}=\sum_l a_{il}a_{lj}, \qquad a_{il}={\lambda_i \lambda_l \over \zeta_i+\zeta_l}, \qquad \lambda_k=\sqrt{c_k(0)} e^{i \zeta_k \bar{z}-i z / 4 \zeta_k}.
\end{equation}
Here $c_k(0)$ and $\zeta_k$ are two sets of constants related to the initial positions and momenta of the N solitons.

This soliton dynamics can be summarized by an associated N-body dynamical system. As for the sine-Gordon solitons \cite{Bowtell} we can deduce the dynamics of sinh-Gordon solitons by following the poles of the Hamiltonian density. Here we denote the sinh-Gordon field as $\phi$ and use the variables $t,x$, the general solution to the sinh-Gordon equation can be written in the form
\begin{equation}
\phi=\ln(f/g)^2.
\end{equation}
Plugging the above ansatz into the Hamiltonian density, we find
\begin{equation}
{\cal H}={2 \over f^2 g^2}\Bigl[(f g_x-g f_x)^2+(f g_t-g f_t)^2+{1 \over 4}(f^2-g^2)^2\Bigr].
\end{equation}
One has the poles of the Hamiltonian density located at
\begin{equation}
f g=0.
\label{pole}
\end{equation}
Firstly, let us consider the one-soliton solution to the sinh-Gordon equation
\begin{equation}
\phi_{s,\bar{s}}=\pm \ln\Bigl[\tanh{(x-x_0)-v t \over 2\sqrt{1-v^2}}\Bigr]^2,
\end{equation}
where $x_0$ is the initial position of the soliton and $v$ is the velocity of the soliton. The motion of poles are easily determined by (\ref{pole}) and we get $x(t)=x_0+vt$, which represents a free motion of the pole. The rest mass of the soliton diverges at $x=0$. However, in this whole analysis of dynamics, the rest mass turns out to be an overall multiplier and we can set $m=1$.

Secondly, for the two-soliton solution in the center-of-mass frame
\begin{equation}
\phi_{ss}=\ln \Bigl[ {v\cosh(\gamma x)-\cosh(\gamma v t) \over v\cosh(\gamma x)+\cosh(\gamma v t)} \Bigr]^2,
\end{equation}
where $\gamma=(1-v^2)^{-1/2}$ and $v$ is the relative velocity of the two solitons, the poles are located at
\begin{equation}
fg=v^2\cosh^2(\gamma x)-\cosh^2(\gamma v t)=0,
\end{equation}
so that the trajectories are
\begin{equation}
x(t)=\pm {1 \over \gamma} \cosh^{-1} \Bigl[ {1 \over v} \cosh (\gamma v t) \Bigr].
\end{equation}
These motions are the same as the sine-Gordon solitons \cite{Bowtell}. Similarly, the time delay of two-soliton scattering can be easily worked out as
\begin{equation}
\Delta t=\lim_{L \rightarrow \infty}\left({1 \over \gamma v} \cosh^{-1}[v \cosh (\gamma x)] |_{-L}^L-{2 L \over v} \right)={2 \over \gamma v}\ln v,
\label{timedelay}
\end{equation}
which of course agrees with the sine-Gordon soliton-soliton scattering.

For completeness, we also list the soliton-antisoliton scattering solution
\begin{equation}
\phi_{s\bar{s}}=\ln \Bigl[ {v\sinh(\gamma x)-\sinh(\gamma v t) \over v\sinh(\gamma x)+\sinh(\gamma v t)} \Bigr]^2,
\end{equation}
in which case the trajectories of the poles are
\begin{equation}
x(t)=\pm {1 \over \gamma} \sinh^{-1} \Bigl[ {1 \over v} \sinh (\gamma v t) \Bigr].
\end{equation}
It has the same time delay (\ref{timedelay}) for the soliton-antisoliton scattering. 

We see that the trajectories and time delays are the same with those of the sine-Gordon theory. The dynamics therefore can be summarized by a N-body Hamiltonian of Ruijsenaars and Schneider \cite{Ruijsenaars} form
\begin{equation}
H=\sum_{j=1}^N \cosh \theta_j \prod_{k \neq j} f(q_j-q_k),
\end{equation}
where $\theta_j$ is the rapidity of particle $j$ and $q_j$ is the canonically conjugate position. The Ruijsenaars-Schneider Hamiltonian for two particles in the center-of-mass frame is given by 
\begin{equation}
H=\cosh\theta~W(q),
\end{equation}
where $\theta$ is the center-of-mass rapidity and its conjugate $q$ measures the interparticle displacement. The potentials are given by
\begin{equation}
W_r(q)=\left| \coth \left({q \over 2}\right) \right|,\quad W_a(q)=\left| \tanh \left({q \over 2}\right) \right|,
\end{equation}
for soliton-soliton scattering and soliton-antilsoliton scattering, respectively.

\subsection{One-spike solution}

The inverse scattering method gives us a procedure to generate string solutions from those of the sinh-Gordon. Essentially the sinh-Gordon configuration serves as a (time-dependent) potential in the lax scattering equations. The string coordinates are then directly obtained from the corresponding wavefunctions. We have in our previous work seen a one-to-one correspondence between the solitons/singularities of the sinh-Gordon field and spikes of the string solution.

Setting N=1 in the spinor (\ref{eqn327}, \ref{eqn328}), we have
\begin{eqnarray}
\varphi_1&=&{c_1(0)(2\zeta_1)^2 e^{2i\zeta_1 \bar{z}+i z /2\zeta_1} \over (\zeta+\zeta_1) (c_1^2(0) e^{4i\zeta_1 \bar{z}}-4\zeta_1^2 e^{i z/ \zeta_1})} e^{i \zeta \bar{z}-i z / 4\zeta}, \\
\varphi_2&=&\left[1-{c_1^2(0)(2\zeta_1) e^{4i\zeta_1 \bar{z}} \over (\zeta+\zeta_1)(c_1^2(0) e^{4i\zeta_1 \bar{z}}-4\zeta_1^2 e^{i z/ \zeta_1})}\right] e^{i \zeta \bar{z}-i z / 4\zeta}.
\end{eqnarray}
Plugging into (\ref{eqn326}), we get the sinh-Gordon field
\begin{equation}
\hat{u}(z,\bar{z})=-\sinh^{-1} \left[{8 c_1(0) \zeta_1 (c_1^2(0) e^{4i\zeta_1 \bar{z}}+4\zeta_1^2 e^{i z /\zeta_1}) e^{2i\zeta_1 \bar{z}+i z /2\zeta_1} \over (c_1^2(0) e^{4i\zeta_1 \bar{z}}-4\zeta_1^2 e^{i z /\zeta_1})^2}\right],
\end{equation}
where $c_1(0)$ and $\zeta_1$ are purely imaginary in order to make the field real.

Now we proceed to write down the first spinor $\phi$
\begin{equation}
\phi=g_A^{-1}\varphi, \qquad \hat{u}=-\hat{\alpha}, \qquad \zeta=-\lambda /2,
\end{equation}
and find
\begin{multline}
\phi_1={(1+i) \over 2\sqrt{2}} e^{-{1 \over 2}(i\lambda \bar{z}-i z /\lambda)} \Bigl\{ \sqrt[8]{-v \over u} e^{{1 \over 4}\hat{\alpha}}{(c_1(0)(2\zeta_1+\lambda)e^{2i \zeta_1 \bar{z}}-2\zeta_1(2\zeta_1-\lambda)e^{i z /2 \zeta_1}) \over (2\zeta_1 -\lambda)(c_1(0) e^{2i \zeta_1 \bar{z}}+2\zeta_1 e^{i z /2\zeta_1})} \cr
+\sqrt[8]{u \over -v} e^{-{1 \over 4}\hat{\alpha}}{(c_1(0)(2\zeta_1+\lambda)e^{2i \zeta_1 \bar{z}}+2\zeta_1(2\zeta_1-\lambda)e^{i z /2 \zeta_1}) \over (2\zeta_1 -\lambda)(c_1(0) e^{2i \zeta_1 \bar{z}}-2\zeta_1 e^{i z /2\zeta_1})} \Bigr\},
\end{multline}
\begin{multline}
\phi_2={(1-i) \over 2\sqrt{2}} e^{-{1 \over 2}(i\lambda \bar{z}-i z /\lambda)} \Bigl\{ -\sqrt[8]{-v \over u} e^{{1 \over 4}\hat{\alpha}}{(c_1(0)(2\zeta_1+\lambda)e^{2i \zeta_1 \bar{z}}-2\zeta_1(2\zeta_1-\lambda)e^{i z /2 \zeta_1}) \over (2\zeta_1 -\lambda)(c_1(0) e^{2i \zeta_1 \bar{z}}+2\zeta_1 e^{i z /2\zeta_1})} \cr
+\sqrt[8]{u \over -v} e^{-{1 \over 4}\hat{\alpha}}{(c_1(0)(2\zeta_1+\lambda)e^{2i \zeta_1 \bar{z}}+2\zeta_1(2\zeta_1-\lambda)e^{i z /2 \zeta_1}) \over (2\zeta_1 -\lambda)(c_1(0) e^{2i \zeta_1 \bar{z}}-2\zeta_1 e^{i z /2\zeta_1})} \Bigr\}.
\end{multline}
For real $\lambda$, the components of the spinor $\phi$ are normalized to be
\begin{equation}
\phi_1^* \phi_1-\phi_2^* \phi_2=1.
\end{equation}
Similarly, for the second spinor $\psi$, we find
\begin{eqnarray}
\psi_1&=&{1 \over 2\sqrt{2}}\left[ e^{-{1 \over 2}(\lambda \bar{z}+z /\lambda)} a_1 + e^{{1 \over 2}(\lambda \bar{z}+z /\lambda)} b_1 \right], \\
\psi_2&=&{1 \over 2\sqrt{2}}\left[ e^{-{1 \over 2}(\lambda \bar{z}+z /\lambda)} a_2 - e^{{1 \over 2}(\lambda \bar{z}+z /\lambda)} b_2 \right],
\end{eqnarray}
where
\begin{eqnarray}
a_1 &\equiv& \Bigl\{ \sqrt[8]{-v \over u} e^{{1 \over 4}\hat{\alpha}}{(c_1(0)(2i\zeta_1+\lambda)e^{2i \zeta_1 \bar{z}}-2\zeta_1(2i\zeta_1-\lambda)e^{i z /2 \zeta_1}) \over (2i\zeta_1 -\lambda)(c_1(0) e^{2i \zeta_1 \bar{z}}+2\zeta_1 e^{i z /2\zeta_1})} \cr
& &+i\sqrt[8]{u \over -v} e^{-{1 \over 4}\hat{\alpha}}{(c_1(0)(2i\zeta_1+\lambda)e^{2i \zeta_1 \bar{z}}+2\zeta_1(2i\zeta_1-\lambda)e^{i z /2 \zeta_1}) \over (2i\zeta_1 -\lambda)(c_1(0) e^{2i \zeta_1 \bar{z}}-2\zeta_1 e^{i z /2\zeta_1})} \Bigr\},
\end{eqnarray}
\begin{eqnarray}
b_1 &\equiv& \Bigl\{ i\sqrt[8]{-v \over u} e^{{1 \over 4}\hat{\alpha}}{(c_1(0)(2i\zeta_1-\lambda)e^{2i \zeta_1 \bar{z}}-2\zeta_1(2i\zeta_1+\lambda)e^{i z /2 \zeta_1}) \over (2i\zeta_1 +\lambda)(c_1(0) e^{2i \zeta_1 \bar{z}}+2\zeta_1 e^{i z /2\zeta_1})} \cr
& &+\sqrt[8]{u \over -v} e^{-{1 \over 4}\hat{\alpha}}{(c_1(0)(2i\zeta_1-\lambda)e^{2i \zeta_1 \bar{z}}+2\zeta_1(2i\zeta_1+\lambda)e^{i z /2 \zeta_1}) \over (2i\zeta_1 +\lambda)(c_1(0) e^{2i \zeta_1 \bar{z}}-2\zeta_1 e^{i z /2\zeta_1})} \Bigr\},
\end{eqnarray}
\begin{eqnarray}
a_2 &\equiv& \Bigl\{ i\sqrt[8]{-v \over u} e^{{1 \over 4}\hat{\alpha}}{(c_1(0)(2i\zeta_1+\lambda)e^{2i \zeta_1 \bar{z}}-2\zeta_1(2i\zeta_1-\lambda)e^{i z /2 \zeta_1}) \over (2i\zeta_1 -\lambda)(c_1(0) e^{2i \zeta_1 \bar{z}}+2\zeta_1 e^{i z /2\zeta_1})} \cr
& &+\sqrt[8]{u \over -v} e^{-{1 \over 4}\hat{\alpha}}{(c_1(0)(2i\zeta_1+\lambda)e^{2i \zeta_1 \bar{z}}+2\zeta_1(2i\zeta_1-\lambda)e^{i z /2 \zeta_1}) \over (2i\zeta_1 -\lambda)(c_1(0) e^{2i \zeta_1 \bar{z}}-2\zeta_1 e^{i z /2\zeta_1})} \Bigr\},
\end{eqnarray}
\begin{eqnarray}
b_2 &\equiv& \Bigl\{ \sqrt[8]{-v \over u} e^{{1 \over 4}\hat{\alpha}}{(c_1(0)(2i\zeta_1-\lambda)e^{2i \zeta_1 \bar{z}}-2\zeta_1(2i\zeta_1+\lambda)e^{i z /2 \zeta_1}) \over (2i\zeta_1 +\lambda)(c_1(0) e^{2i \zeta_1 \bar{z}}+2\zeta_1 e^{i z /2\zeta_1})} \cr
& &+i\sqrt[8]{u \over -v} e^{-{1 \over 4}\hat{\alpha}}{(c_1(0)(2i\zeta_1-\lambda)e^{2i \zeta_1 \bar{z}}+2\zeta_1(2i\zeta_1+\lambda)e^{i z /2 \zeta_1}) \over (2i\zeta_1 +\lambda)(c_1(0) e^{2i \zeta_1 \bar{z}}-2\zeta_1 e^{i z /2\zeta_1})} \Bigr\}.
\end{eqnarray}
For real $\lambda$, the components of the spinor $\psi$ are normalized to be
\begin{equation}
\psi_1^* \psi_1-\psi_2^* \psi_2=1.
\end{equation}
In order for a simpler expression, the string solution is presented as
\begin{eqnarray}
Z_1 &\equiv& Y_{-1}+i Y_0=\phi_1^* \psi_1-\phi_2^* \psi_2, \\
Z_2 &\equiv& Y_1+i Y_2=\phi_2^* \psi_1^*-\phi_1^* \psi_2^*.
\end{eqnarray}
Recalling the change of variables (\ref{eqn39}), the one-spike string solution is then found to be
\begin{multline}
Z_1={e^{{1+i \over 2}(i\lambda\bar{z}'-z'/\lambda)} \over 2(c_1(0) e^{2i\zeta_1\bar{z}'}-2\zeta_1 e^{i z'/2 \zeta_1})}\Bigl\{2\zeta_1 e^{i z'/2 \zeta_1}(1+e^{\lambda\bar{z}'+z'/\lambda}) \cr
+c_1(0) e^{2i\zeta_1 \bar{z}'}{(2\zeta_1-\lambda)((2i\zeta_1+\lambda)^2+e^{\lambda\bar{z}'+z'/\lambda}(2i \zeta_1-\lambda)^2) \over (2\zeta_1+\lambda)(4\zeta_1^2+\lambda^2)} \Bigr\},
\end{multline}
\begin{multline}
Z_2={i e^{{1+i \over 2}(i\lambda\bar{z}'-z'/\lambda)} \over 2(c_1(0) e^{2i\zeta_1\bar{z}'}-2\zeta_1 e^{i z'/2 \zeta_1})}\Bigl\{2\zeta_1 e^{i z'/2 \zeta_1}(1-e^{\lambda\bar{z}'+z'/\lambda}) \cr
+c_1(0) e^{2i\zeta_1 \bar{z}'}{(2\zeta_1-\lambda)((2i\zeta_1+\lambda)^2-e^{\lambda\bar{z}'+z'/\lambda}(2i \zeta_1-\lambda)^2) \over (2\zeta_1+\lambda)(4\zeta_1^2+\lambda^2)} \Bigr\},
\end{multline}
It is interesting to note that $u(\bar{z})$ and $v(z)$ only come into $\bar{z}'$ and $z'$, respectively. This is the residual conformal symmetry which can be further used to fix the time-like conformal gauge.

\subsection{N-spike solution}

Now we consider the general sinh-Gordon solution with N solitons. The first spinor is solved to be
\begin{eqnarray}
\phi_1&=&-{(1+i) \over 2\sqrt{2}}e^{-{1 \over 2}(i\lambda \bar{z}-i z /\lambda)}\Bigl\{ \sqrt[8]{-v \over u} e^{{1 \over 4}\hat{\alpha}}(\tilde{\varphi}_2-\tilde{\varphi}_1)_-^1 +\sqrt[8]{u \over -v} e^{-{1 \over 4}\hat{\alpha}}(\tilde{\varphi}_2+\tilde{\varphi}_1)_-^1 \Bigr\}, \\
\phi_2&=&+{(1-i) \over 2\sqrt{2}}e^{-{1 \over 2}(i\lambda \bar{z}-i z /\lambda)}\Bigl\{ \sqrt[8]{-v \over u} e^{{1 \over 4}\hat{\alpha}}(\tilde{\varphi}_2-\tilde{\varphi}_1)_-^1 -\sqrt[8]{u \over -v} e^{-{1 \over 4}\hat{\alpha}}(\tilde{\varphi}_2+\tilde{\varphi}_1)_-^1 \Bigr\},
\end{eqnarray}
where
\begin{equation}
(\tilde{\varphi}_2 \pm \tilde{\varphi}_1)_\pm^1=1 \pm \sum_{j,l}{\lambda_j \over \pm {\lambda \over 2}+\zeta_j}(1-A)_{jl}^{-1}\lambda_l+\sum_{j,l,k}{\lambda_j \over \pm {\lambda \over 2}+\zeta_j}{\lambda_j \lambda_l \over \zeta_j+\zeta_l}(1-A)_{lk}^{-1}\lambda_k. \label{eqn363}
\end{equation}
The subscript $\pm$ corresponds to the $\pm$ before the spectral parameter $\lambda$. The second spinor is solved to be
\begin{multline}
\psi_1=-{1 \over 2\sqrt{2}}\Bigl\{\sqrt[8]{-v \over u} e^{{1 \over 4}\hat{\alpha}}\Bigl[e^{-{1 \over 2}(\lambda\bar{z}+z/\lambda)}(\tilde{\varphi}_2-\tilde{\varphi}_1)_+^2 +i e^{{1 \over 2}(\lambda\bar{z}+z/\lambda)}(\tilde{\varphi}_2-\tilde{\varphi}_1)_-^2\Bigr] \cr
+i\sqrt[8]{u \over -v} e^{-{1 \over 4}\hat{\alpha}}\Bigl[e^{-{1 \over 2}(\lambda\bar{z}+z/\lambda)}(\tilde{\varphi}_2+\tilde{\varphi}_1)_+^2 -i e^{{1 \over 2}(\lambda\bar{z}+z/\lambda)}(\tilde{\varphi}_2+\tilde{\varphi}_1)_-^2 \Bigr],
\end{multline}
\begin{multline}
\psi_2=-{i \over 2\sqrt{2}}\Bigl\{\sqrt[8]{-v \over u} e^{{1 \over 4}\hat{\alpha}}\Bigl[e^{-{1 \over 2}(\lambda\bar{z}+z/\lambda)}(\tilde{\varphi}_2-\tilde{\varphi}_1)_+^2 +i e^{{1 \over 2}(\lambda\bar{z}+z/\lambda)}(\tilde{\varphi}_2-\tilde{\varphi}_1)_-^2\Bigr] \cr
-i\sqrt[8]{u \over -v} e^{-{1 \over 4}\hat{\alpha}}\Bigl[e^{-{1 \over 2}(\lambda\bar{z}+z/\lambda)}(\tilde{\varphi}_2+\tilde{\varphi}_1)_+^2 -i e^{{1 \over 2}(\lambda\bar{z}+z/\lambda)}(\tilde{\varphi}_2+\tilde{\varphi}_1)_-^2 \Bigr],
\end{multline}
where
\begin{equation}
(\tilde{\varphi}_2 \pm \tilde{\varphi}_1)_\pm^2=1\pm \sum_{j,l}{\lambda_j \over \pm{i\lambda \over 2}+\zeta_j}(1-A)_{jl}^{-1}\lambda_l+\sum_{j,l,k}{\lambda_j \over \pm{i\lambda \over 2}+\zeta_j}{\lambda_j \lambda_l \over \zeta_j+\zeta_l}(1-A)_{lk}^{-1}\lambda_k.
\end{equation}
Similar to (\ref{eqn363}), the subscript $\pm$ corresponds to the $\pm$ before the spectral parameter $\lambda$. Recalling the change of variables (\ref{eqn39}), the N-spike string solution is given by
\begin{multline}
Z_1={1-i \over 4}e^{{1 \over 2}(i\lambda\bar{z}'-i z'/\lambda)}\Bigl\{i(\tilde{\varphi}_2-\tilde{\varphi}_1)_+^1 \Bigl[e^{-{1 \over 2}(\lambda\bar{z}'+z'/\lambda)}(\tilde{\varphi}_2+\tilde{\varphi}_1)_+^2 -ie^{{1 \over 2}(\lambda\bar{z}'+z'/\lambda)}(\tilde{\varphi}_2+\tilde{\varphi}_1)_-^2 \Bigr] \cr
+(\tilde{\varphi}_2+\tilde{\varphi}_1)_+^1 \Bigl[e^{-{1 \over 2}(\lambda\bar{z}'+z'/\lambda)}(\tilde{\varphi}_2-\tilde{\varphi}_1)_+^2 +ie^{{1 \over 2}(\lambda\bar{z}'+z'/\lambda)}(\tilde{\varphi}_2-\tilde{\varphi}_1)_-^2 \Bigr] \Bigr\},
\label{eqn250}
\end{multline}
\begin{multline}
Z_2={1+i \over 4}e^{{1 \over 2}(i\lambda\bar{z}'-i z'/\lambda)}\Bigl\{i(\tilde{\varphi}_2-\tilde{\varphi}_1)_+^1 \Bigl[e^{-{1 \over 2}(\lambda\bar{z}'+z'/\lambda)}(\tilde{\varphi}_2+\tilde{\varphi}_1)_+^2 +ie^{{1 \over 2}(\lambda\bar{z}'+z'/\lambda)}(\tilde{\varphi}_2+\tilde{\varphi}_1)_-^2 \Bigr] \cr
+(\tilde{\varphi}_2+\tilde{\varphi}_1)_+^1 \Bigl[e^{-{1 \over 2}(\lambda\bar{z}'+z'/\lambda)}(\tilde{\varphi}_2-\tilde{\varphi}_1)_+^2 -ie^{{1 \over 2}(\lambda\bar{z}'+z'/\lambda)}(\tilde{\varphi}_2-\tilde{\varphi}_1)_-^2 \Bigr] \Bigr\}.
\label{eqn251}
\end{multline}

Let us summarize the properties of the general solution we have constructed. The string solution given above was found on the infinite line, with generalization to the compact circle (with periodic boundary conditions) remaining. The general string solution is characterized by two arbitrary functions $u(\bar{z}),v(z)$ and the discrete set of moduli representing the soliton singularities (coordinates). After fixing the conformal frame only the soliton moduli remain giving a specification of the string moduli.
 
In general the sinh-Gordon singularities behave as particles and follow interacting particle trajectories. Through our explicit transformations this dynamics translates into the spike dynamics of the $AdS_3$ string. Concretely, given the trajectories of N solitons $x_i(t),i=1,2,\cdots,N$ (which as we have shown are governed by a dynamical system of RS type), we can in principle by direct substitution (\ref{eqn250}, \ref{eqn251}) with $\sigma_i(\tau)$ construct the trajectories of N spikes by
\begin{equation}
Z_1^i(\tau)=Z_1(\tau,\sigma_i(\tau)), \qquad Z_2^i(\tau)=Z_2(\tau,\sigma_i(\tau)),
\end{equation}
where $\tau$ acts like the proper time. We therefore have a mapping where on the left hand side the index $i$ labels the string spikes while on the right side it denotes the solitons/singularities. This construction is straightforward in principle, with the map provided by the known wavefunctions of the scattering problem. In explicit terms however it leads to complex expressions. 

Some general features of the dynamics that emerge from the construction can be deduced. First of all the N-body field theory dynamics being integrable it is automatic that the corresponding string theory system defined by our inverse scattering map is also integrable. The soliton N-body interactions have the characteristic that they are of Calogero type (as compared to the Toda, nearest neighbor interactions). It is not obvious, and remains to be established which of these two possible integrable schemes are associated with the spike dynamics of closed AdS strings. Here we also have the analogy and lessons from the recent study of N-body description of magnons on $R \times S^2$. In the magnon case the map between the soliton dynamical system and the `string' system was established in \cite{Aniceto:2008pc}. It involves the multi-Hamiltonian and multi-Poisson structures of the integrable N-body Ruijsenaars-Schneider system. Our present construction implies that such a correspondence is also expected to hold in the AdS case.

\section{Conclusion}

We have in the present paper discussed some solvable aspects of classical string dynamics in $AdS_3$. Using a relationship with the sinh-Gordon theory we have derived a general class of solutions defined on the infinite line. This class of solutions represents a map (to the AdS space) of N-soliton configurations of the sinh-Gordon field theory. 

One notes that the field theory solutions are characterized by singularities which we argue play a central role in string dynamics. We show that the earlier introduced correspondence between singularities of the field configuration and spikes of the string continues to hold in the general case. The singularities of the field theory follow particle-like trajectories and can be described by a N-body problem with nontrivial interactions. We review this dynamics in some detail both in the sinh-Gordon case and the better known limiting case given by the Liouville theory. Through the inverse scattering map the particle picture of field theory is seen to induce a point-like dynamics for the string. This dynamics is seen to concentrate precisely at the locations of the string theory `spikes', the map being between soliton collective coordinates (positions and momenta) and the spike coordinates.

A different attempt to generalize Kruczenski's static solution was presented recently in \cite{Dorey:2008vp} through `gluing'.  In this construction, which is based on a static picture, the equations of motion appear not to be satisfied at the locations of spikes. As we have argued, it is there that the nontrivial dynamics takes place, being determined by the soliton moduli space dynamics. 

Our map between the solitons and string moduli is nonlinear, determined by the inverse scattering construction itself. 
It follows in principle that this dynamical system is integrable, belonging to the same Ruijseenaars-Schneider hierarchy which 
governs the sinh-Gordon theory. It will be interesting to compare this dynamical system with a system proposed by Kruczenski
in \cite{Kruczenski:2004wg}. A relevant continuation of the present work is therefore a generalization (and construction of N-soliton solutions) in the compact case with solitons on the circle. On the field theory side a construction of solutions is known, and the inverse scattering map can be used, similarly to our present discussion, to determine the corresponding AdS string solutions.

\acknowledgments

We would like to thank I. Aniceto, J. Avan, C. Kalousios, M. Spradlin and A. Volovich for comments and discussions. This work is supported by the Department of Energy under contract DE-FG02-91ER40688.

\appendix

\section{Inverse scattering and the spinors}

In this appendix, we summarize the inverse scattering method \cite{AKNS} to solve the Lax pair equations (\ref{laxeqn}) with the matrices (\ref{lax}). Choose the boundary condition of the field as $\hat{u},\bar{\partial}\hat{u} \rightarrow 0$ as $\bar{z} \rightarrow \pm \infty$. The spinor $\varphi$ can be written as an integral form
\begin{equation}
\varphi(\zeta,\bar{z})=\begin{pmatrix} 0 \cr 1 \end{pmatrix} e^{i \zeta \bar{z}}+\int_{\bar{z}}^\infty K(\bar{z},s)e^{i\zeta s} ds ,
\end{equation}
where the kernel satisfies the GLM equation
\begin{equation}
\bar{K}(\bar{z},y)+\begin{pmatrix} 0 \cr 1 \end{pmatrix} F(\bar{z}+y)+\int_{\bar{z}}^\infty K(\bar{z},s)F(s+y) ds=0 ,
\label{glm}
\end{equation}
with the function $F(x)$ defined to be
\begin{equation}
F(x)={1 \over 2\pi}\int_{-\infty}^{\infty} r(k) e^{i k x} dk-i\sum_{j=1}^N c_j e^{i \zeta_j x} ,
\end{equation}
where $r(k)$ is the reflection coefficient, $c_j$ and $\zeta_j$ are constants. In the case of sinh-Gordon with real field $\hat{u}$, the kernels are
\begin{equation}
K=\begin{pmatrix} K_1(\bar{z},s) \cr K_2(\bar{z},s) \end{pmatrix}, \qquad \bar{K}=\begin{pmatrix} K_2(\bar{z},s) \cr K_1(\bar{z},s) \end{pmatrix},
\end{equation}
with $K_1$ and $K_2$ real.

Now we try to solve the GLM equation (\ref{glm}) by some ansatz. Consider the soliton solutions to the sinh-Gordon equation which has $r(k)=0$ and plug in the ansatz
\begin{equation}
K_i(\bar{z},s)=\sum_{j=1}^N \sqrt{c_j} f_{ij}(\bar{z}) e^{i \zeta_j s}, \qquad i=1,2,
\end{equation}
we get
\begin{eqnarray}
f_{1j}&=&i\sum_{k=1}^N (1-A)_{jk}^{-1} \lambda_k , \\
f_{2j}&=&-i\sum_{l,k=1}^N {\lambda_j \lambda_l \over \zeta_j+\zeta_l} (1-A)_{lk}^{-1} \lambda_k ,
\end{eqnarray}
where the matrix $A$ is defined as
\begin{equation}
A_{ij}=\sum_l a_{il}a_{lj}, \qquad a_{il}={\lambda_i \lambda_l \over \zeta_i+\zeta_l}, \qquad \lambda_k=\sqrt{c_k} e^{i \zeta_k \bar{z}}.
\end{equation}
The wavefunctions are solved to be
\begin{eqnarray}
\varphi_1(\zeta,\bar{z})&=&-\Bigl(\sum_{j,l} {\lambda_j \over \zeta+\zeta_j}(1-A)_{jl}^{-1}\lambda_l\Bigr)e^{i \zeta \bar{z}}, \\
\varphi_2(\zeta,\bar{z})&=&\Bigl(1+\sum_{j,l,k} {\lambda_j \over \zeta+\zeta_j}{\lambda_j \lambda_l \over \zeta_j+\zeta_l}(1-A)_{lk}^{-1} \lambda_k \Bigr)e^{i \zeta \bar{z}}.
\end{eqnarray}
Adding the $z$ dependence, we get
\begin{eqnarray}
\varphi_1(\zeta,z,\bar{z})&=&-\Bigl(\sum_{j,l} {\lambda_j \over \zeta+\zeta_j}(1-A)_{jl}^{-1}\lambda_l\Bigr)e^{i \zeta \bar{z}-i z / 4\zeta}, \\
\varphi_2(\zeta,z,\bar{z})&=&\Bigl(1+\sum_{j,l,k} {\lambda_j \over \zeta+\zeta_j}{\lambda_j \lambda_l \over \zeta_j+\zeta_l}(1-A)_{lk}^{-1} \lambda_k \Bigr)e^{i \zeta \bar{z}-i z / 4\zeta},
\end{eqnarray}
with $c_j(z)=c_j(0) e^{-i z / 2\zeta_j}$. The sinh-Gordon field $\hat{u}(z,\bar{z})$ is found to be
\begin{equation}
\hat{u}(z,\bar{z})=\sinh^{-1} \Bigl[{4\zeta \over i}{\partial(\varphi_1 \varphi_2) \over (\varphi_1)^2-(\varphi_2)^2}\Bigr].
\end{equation}

\section{More spiky strings in flat spacetime}

In this appendix, we discuss the more involved case with two spikes, i.e., $N=N_A=2,N_B=1$. The arbitrary functions are
\begin{equation}
f(\sigma^+)={c_1 \over y_1-\sigma^+}+{c_2 \over y_2-\sigma^+}, \qquad g(\sigma^-)={d_1 \over z_1-\sigma^-}.
\end{equation}
The expressions for the constants $c_i,y_i$ are already very complicated. For simplicity, using the initial data (\ref{inidata}) and setting $v_1=v_0,v_2=-v_0,\sigma_1^0=-\sigma_0,\sigma_2^0=\sigma_0$, we have the trajectories of singularities
\begin{equation}
\sigma_{1,2}(\tau)=\mp \sqrt{\tau^2-2\tau v_0 \sigma_0+\sigma_0^2}.
\end{equation}
After some calculation, the string solution is found to be
\begin{multline}
X^0={u \over 6\sqrt{2}d_1(1-v_0^2)\sigma_0^2}\Bigl(2(z_1-v_0\sigma_0)^2 (\tilde{\sigma}^+)^3+d_1^2(\sigma^+-v_0\sigma_0)^3\Bigr) \cr
+{v \over \sqrt{2}d_1}\Bigl({1 \over 3}(\sigma^- - z_1)^3+{1 \over 2}d_1^2 \sigma^- \Bigr),
\end{multline}
\begin{multline}
X^1={u \over 6\sqrt{2}d_1(1-v_0^2)\sigma_0^2}\Bigl(2(z_1-v_0\sigma_0)^2 (\tilde{\sigma}^+)^3-d_1^2(\sigma^+-v_0\sigma_0)^3\Bigr) \cr
+{v \over \sqrt{2}d_1}\Bigl({1 \over 3}(\sigma^- - z_1)^3-{1 \over 2}d_1^2 \sigma^- \Bigr),
\end{multline}
\begin{multline}
X^2={u(z_1 - v_0 \sigma_0) \over (1-v_0^2)\sigma_0^2}\Bigl( {1 \over 3}(\sigma^+)^3+{\sigma_0^2-2 z_1 v_0 \sigma_0 + v_0^2 \sigma_0^2 \over 2(z_1-v_0 \sigma_0)} (\sigma^+)^2 -v_0 \sigma_0^2 {\sigma_0-v_0 z_1 \over z_1-v_0 \sigma_0} \sigma^+ \Bigr) \cr
+{v \over 2}(\sigma^- - z_1)^2,
\end{multline}
where
\begin{equation}
\tilde{\sigma}^+ \equiv \sigma^+ + \sigma_0 {\sigma_0-v_0 z_1 \over z_1-v_0 \sigma_0}.
\end{equation}

\end{document}